\documentclass[a4paper,showpacs,superscriptaddress,twocolumn]{revtex4}
\usepackage[english]{babel}
\usepackage{graphicx}
\usepackage{xspace}
\usepackage{fancybox}
\usepackage{epic}
\usepackage{calc}
\usepackage{ifthen}
\usepackage{float}
\usepackage{color}
\usepackage{amsmath,amssymb}
\usepackage{rotating}
\usepackage{subfigure}
\usepackage{exscale}
\usepackage{setspace}
\usepackage[latin1]{inputenc}
\begin{document}
\title{Diffusion based degradation mechanisms in giant magnetoresistive spin valves}
\author{Matthias Hawraneck}\email[]{matthias.hawraneck@infineon.com}
\affiliation{Department Sense and Control, Infineon Technologies
AG, 85579 Neubiberg, Germany} \affiliation{Institute for Materials
Science, Technische Universität Darmstadt, 64287 Darmstadt,
Germany}
\author{Jürgen Zimmer}
\affiliation{Department Sense and Control, Infineon Technologies
AG, 85579 Neubiberg, Germany}
\author{Wolfgang Raberg}
\affiliation{Department Sense and Control, Infineon Technologies
AG, 85579 Neubiberg, Germany}
\author{Klemens Prügl}
\affiliation{Department Sense and Control, Infineon Technologies
AG, 85579 Neubiberg, Germany}
\author{Stephan Schmitt}
\affiliation{Department Sense and Control, Infineon Technologies
AG, 85579 Neubiberg, Germany}
\author{Thomas Bever}
\affiliation{Department Sense and Control, Infineon Technologies
AG, 85579 Neubiberg, Germany}
\author{Stefan Flege}
\affiliation{Institute for Materials Science, Technische
Universität Darmstadt, 64287 Darmstadt, Germany}
\author{Lambert Alff}
\affiliation{Institute for Materials Science, Technische
Universität Darmstadt, 64287 Darmstadt, Germany}

\date{23 April 2008}

\begin{abstract}
Spin valve systems based on the giant magnetoresistive (GMR)
effect as used for example in hard disks and automotive
applications consist of several functional metallic thin film
layers. We have identified by secondary ion mass spectrometry
(SIMS) two main degradation mechanisms: One is related to oxygen
diffusion through a protective cap layer, and the other one is
interdiffusion directly at the functional layers of the GMR stack.
By choosing a suitable material as cap layer (TaN), the oxidation
effect can be suppressed.

\end{abstract}

\maketitle

%%%%%%%%%%%%%%%%%%%%
%%% INTRODUCTION %%%
%%%%%%%%%%%%%%%%%%%%

Due to its huge application potential the giant magnetoresistance
(GMR) effect has aroused much public interest after its discovery
in 1988 by last year's nobel prize winners A.~Fert and P.~Grünberg
\cite{Baibich:88,Binasch:89}. Subsequently, Dieny {\em et al.}
have demonstrated the spin-valve effect in not coupled magnetic
layers \cite{Dieny:91}. Besides the well-known use of spin-valves
in read heads of hard disk drives, they can also be effectively
used in the automotive industry to detect rotational speed,
direction or angle in a variety of applications. The required
lifetime and the temperature profile in these kind of applications
demand an increased thermal stability, in particular in safety
relevant parts. A GMR device generally consists of several
different metallic layers. In our case we are dealing with a
so-called bottom-pinned spin valve (BSV) where the bottom
electrode is magnetically fixed by the exchange bias effect
\cite{Nogues:99}. The use of such devices in industrial
environments and cars defines restrictions with respect to thermal
stability. In previous studies of thermal stability of spin valve
structures, several groups have focused on diffusion effects
\cite{Anderson:00,Kim:01,Huang:01}. Cho {\em et al.} have
investigated the use of amorphous CoNbZr as seed and capping layer
\cite{Cho:02}, while Kim {\em et al.} have focused on Mn diffusion
\cite{Kim:06}.

%absatz
In our experiments we investigated GMR stacks with Ta and TaN cap
layers with 5 resp.~10\,nm thickness, and stored them at
300$^{\circ}$C in ambient atmosphere. After defined time steps we
measured the magnetic and electrical properties of the stored
samples. To evaluate the depth dependent composition changes due
to diffusion effects in the layered structure, we performed
secondary ion mass spectrometry (SIMS) measurements.

%%%%%%%%%%%%%%%%%%%%
%%% EXPERIMENTAL %%%
%%%%%%%%%%%%%%%%%%%%

The deposition of the bottom-pinned spin-valves (BSV) is performed
in a commercial sputter system on $8^{\prime\prime}$-wafers. The
stack system comprises a Ta and NiFe seed layer, PtMn as the
natural antiferromagnet, a CoFe pinned layer, a Ru spacer layer, a
CoFe reference layer, the Cu spacer layer and a CoFe free (sense)
layer. The composition of CoFe is 90\% Co and 10\% Fe for all CoFe
layers. The cap layer material is Ta or TaN, and the thickness of
the cap layer is 5\,nm or 10\,nm, respectively. The wafers are
annealed at 280$^{\circ}$C in a high magnetic field of 1\,T to
change the PtMn from the paramagnetic fcc to the antiferromagnetic
$L1_0$ phase \cite{Lederman:99}. This annealing leads to exchange
coupling between the natural antiferromagnet PtMn and the
reference layer. The storage experiment itself was done with small
pieces of the original wafer having a size of 12x10\,mm$^2$. To
avoid random effects, e.g.~the influence of particles, we always
processed three pieces of each wafer. Before storage and after
defined storage intervals of 5\,h, 10\,h, 20\,h and 50\,h in
ambient atmosphere at 300$^\circ$C, we investigated the sample
properties by several methods. The sheet resistance in dependence
of an external field was measured using a probe head for a 4 point
measurement (Jandel). The magnetic field range was $\pm$40\,mT.
Depth-dependent composition analysis was performed by SIMS (Cameca
ims5f) with 1.3\,keV O$_2^+$ and 5.5\,keV Ar$^+$ as well as Cs$^+$
ions detecting positive secondary ions.

\begin{figure}[b]
\centering{%
\includegraphics[width=0.85\columnwidth,clip=]{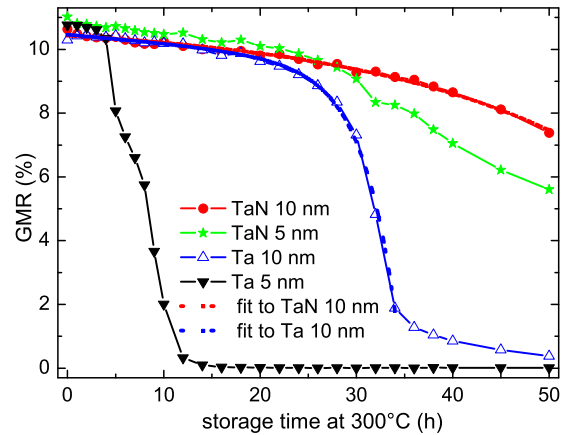}}
\caption{Decay of GMR effect in stacks with Ta and TaN cap layers
of different thickness as a function of storage time at
300$^\circ$C. The two fit curves for TaN 10\,nm (1 decay mode) and
Ta 5\,nm (2 decay modes) can hardly be distinguished from the
experimental data.}\label{Fig:GMR}
\end{figure}

\begin{figure}[t]
\centering{%
\includegraphics[width=0.85\columnwidth,clip=]{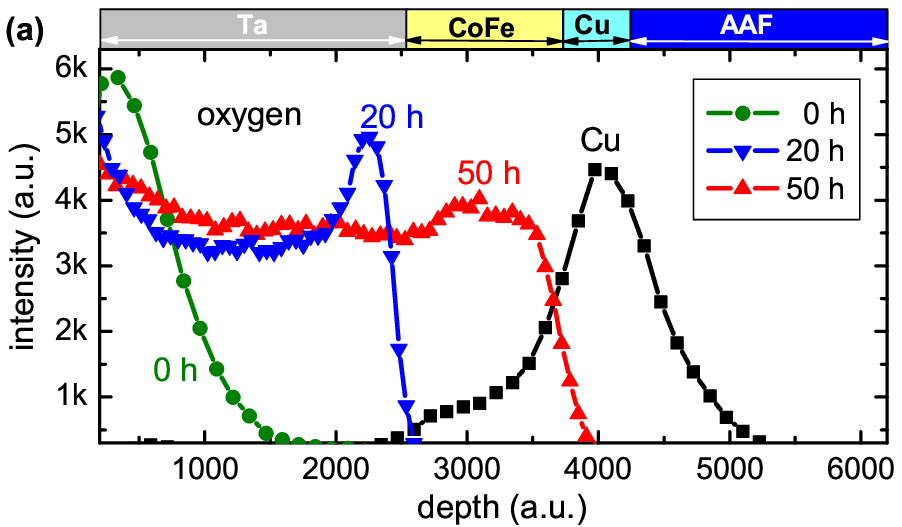}
\includegraphics[width=0.85\columnwidth,clip=]{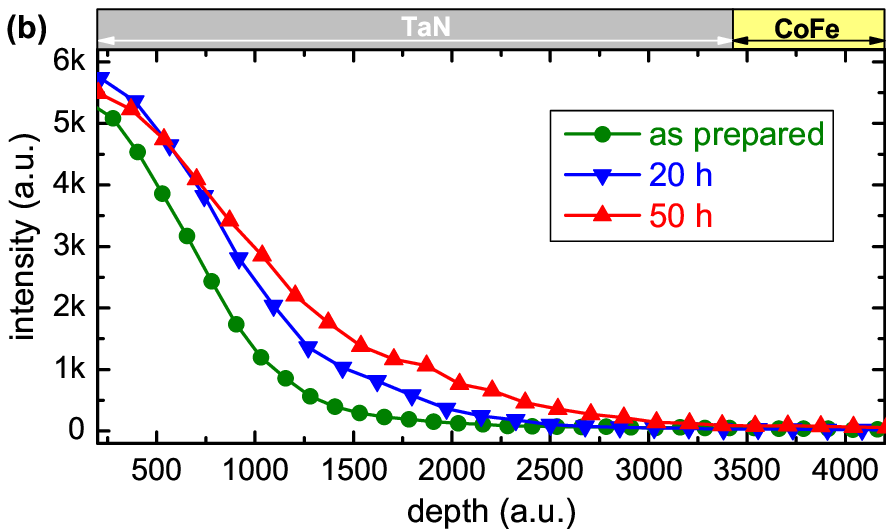}
\includegraphics[width=0.85\columnwidth,clip=]{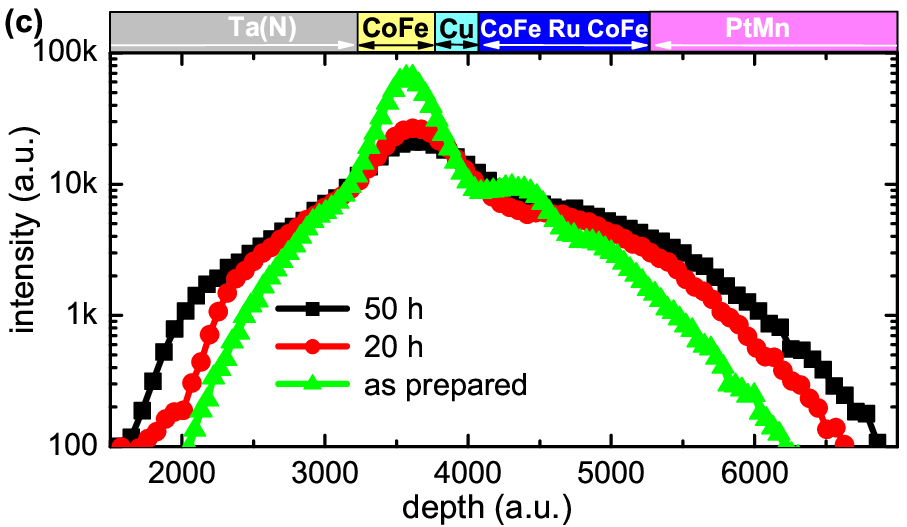}}
\caption{Secondary ion mass spectrometry profiles of spin valves
with different cap layers and after different storage times at
300$^\circ$C. (a) Oxygen signal for a BSV with 10\,nm Ta cap, Cu
peak as reference. (b) Oxygen signal for a BSV with 10\,nm TaN
cap. (c) Cobalt signal for a BSV with 5\,nm TaN cap. The
corresponding layers within the GMR stacks are indicated above
each panel.}\label{Fig:SIMS}
\end{figure}

%%%%%%%%%%%%%%%%%%%%%%%%%%%%
%%% DESCRIPTION FIGURE 1 %%%
%%%%%%%%%%%%%%%%%%%%%%%%%%%%

Fig.~1 shows the GMR effect of spin valves which differ only in
the cap layer material as a function of the thermal stress time.
The initial GMR effect differs slightly from sample to sample
between 10.2 and 11.0\,\% due to intrinsic process variations. The
change of GMR effect in the first 4 hours of thermal stress at
300\,°C is very similar for all samples, but after 5 hours the
stack with the 5\,nm Ta cap shows a strong degradation which ends
in a complete loss of GMR effect after 14\,h. The stack with a Ta
cap layer of double thickness (10\,nm) undergoes the same fate,
but only after 30\,h. In contrast to the Ta capped stacks, the
stacks with a 5 or 10\,nm thick TaN cap layer do not show this
sudden decrease in GMR, but a continuous decrease over time. Note
that this observed decrease is acceptable for automotive
applications.

A similar picture is observed for the change in stack resistance.
The initial sheet resistance varies between 15.9\,$\Omega/\square$
and 17.2\,$\Omega/\square$. After 8\,h of storage at 300$^\circ$C
the resistance of the stack with the 5\,nm Ta cap increases
sharply. This increase already decelerates after 12\,h, but even
after 50\,h of storage no saturation is observed. The resistance
increases to $\sim 60\,\Omega/\square$, which corresponds to an
increase of more than 250\,\%. In contrast, the resistance changes
for the other stacks are comparably small in the range of 5\,\%
resp.~25\,\%.

Also shown in Fig.~1 are exponential decay fits to the curves of
the 10\,nm TaN and Ta capped stacks. While the decay in the case
of the TaN cap layer can be described with one decay mode, for the
Ta only cap layer, two decay modes fit better the experimental
data. This is a first hint, that two different degradation
mechanisms are at work in the GMR stacks.

%%%%%%%%%%%%%%%%%%%%%%%%%%%%
%%% DESCRIPTION FIGURE 2 %%%
%%%%%%%%%%%%%%%%%%%%%%%%%%%%

To further investigate the origin of the degradation mechanisms
which lead to the experimentally observed behavior, we performed
SIMS experiments.

Fig.~2(a) shows the depth dependent oxygen intensities of a 10\,nm
Ta capped GMR stack after different annealing times. As a
reference the copper signal is shown for an as-prepared sample.
Within the whole figure, the corresponding layers of the GMR stack
are shown above each individual panel. Because of different
sputter rates for oxidized and non-oxidized Ta, we adjusted the
depth for the samples with 20\,h and 50\,h of storage, by
overlapping the copper peaks. This method is valid because the
thickness of the layers and the copper peak distribution is
unaffected by the oxidation.

Already the deposited stack comprises a small oxygen peak which
is, however, clearly separated from the copper peak. The oxygen is
only detectable in the Ta cap. After 20\,h of storage the TaO
layer has grown and the oxygen front reaches the CoFe free layer.
At the same time the decay of the GMR effect is strongly
accelerated (compare Fig.~1). In the next hours of storage the
oxidation proceeds and after 50\,h it reaches the Cu spacer layer.
At this point only 0.2\% of the original GMR effect is detectable.

A different picture is observed for the BSV with a 10\,nm thick
TaN cap (Fig.~2(b)). A distinctive intensity of oxygen signal is
detected after sample preparation in the top of the cap, too, but
the broadening of this signal is very slow. After 50\,h of storage
at 300$^\circ$C, the oxygen content in the area of the free layer
is unchanged. Between 0\,h and 50\,h of storage, we observe a loss
of GMR of 3\% which cannot be attributed to the small increase of
oxygen signal in the cap layer.

The data shown clearly identifies oxidation of the CoFe free layer
as the first degradation mechanism leading to performance failure.
The origin is the diffusion of oxygen through the cap layer. Our
data also clearly shows, how a different choice of material for
the cap layer (TaN) suppresses this mechanism effectively to an
acceptable level for applications.

The continuous decrease of the GMR effect by annealing indicates
that a second mechanism contributes to degradation. In Fig.~2(c)
we also plotted the relative intensity of the cobalt signal in a
logarithmic scale for a stack with a 5\,nm thick TaN cap before
storage and after 20h and 50h of storage. The highest cobalt
intensity is detected in the area of the free layer; the two
additional Co containing layers in the artificial antiferromagnet
can be identified by the shoulders on the line of decreasing
intensity towards higher depth in the case of the as prepared
sample. With increased storage time the Co peaks broaden and the
two CoFe layers in the AAF cannot be resolved anymore. Thus there
is interdiffusion in the region of the functional layers of the
GMR stack. Other components that are not shown in Fig.~2 also
participate in this interdiffusion, e.g.~Mn from the natural
antiferromagnet layer which exhibits significant intensities in
the region between the cap layer and the free layer after 50\,h
storage time. Thus, we have identified interdiffusion as a second
source of degradation.

%%%%%%%%%%%%%%%%%%
%%% DISCUSSION %%%
%%%%%%%%%%%%%%%%%%

\begin{figure}
\centering{%
\includegraphics[width=0.85\columnwidth,clip=]{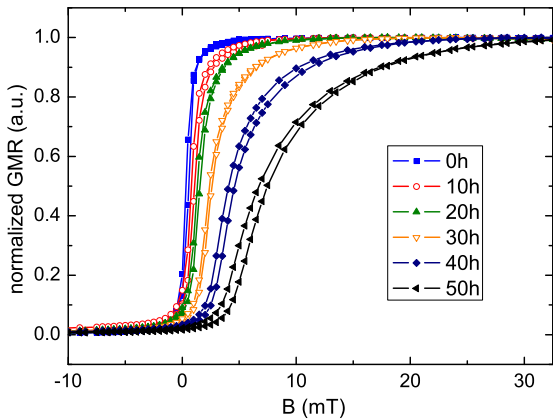}}
\caption{Broadening of the normalized GMR vs applied field for
bottom-pinned spin valves with a 10\,nm TaN cap laver for storage
times from 0\,h to 50\,h at 300$^\circ$C.}\label{Fig:GMRbroad}
\end{figure}

The most relevant functional layers within the GMR stack are the
free layer (CoFe), the spacer (Cu), and the reference layer
(CoFe). It is obvious that oxidation of these layers will
immediately start to decrease and eventually destroy the GMR
effect. The observed oxidation of the free layer results in a
reduction of the spin-diffusion-length and the GMR effect first is
reduced. When the complete free layer is oxidized, spin dependent
scattering vanishes, and hence no GMR effect can be observed. By
choice of a suitable cap layer material as TaN, the oxygen
diffusion and oxidation mechanism is sufficiently suppressed.

The second mechanism of degradation is not related to the oxygen.
The continuous decrease of the GMR effect occurs similarly in all
stacks independent on the choice of cap material or cap layer
thickness. As can be concluded from the SIMS profiles, Co
interdiffusion within the functional region occurs. The
interdiffusion at the functional layers increases the resistance
and surface roughness. Both effects contribute to a reduction of
the GMR effect. The increased interface roughness also leads to a
higher ferromagnetic N\'{e}el-coupling between the free and
reference layer. In turn, this leads to a shift of the minor loop
and the broadening of the hysteresis curve, as is observed for our
devices after high-temperature storage (see Fig.~3)
\cite{Neel:62}. Note, that due to the high activation energy of
$E_{\text{a}}\geq 1.4\pm0.1$\,eV of this process the observed
reduction in GMR by the interdiffusion effect is within the
acceptable limits even for safety relevant applications in the
automotive industry. A further improvement of the device
performance could be obtained by tayloring diffusion barrier
layers into the stack.

%%%%%%%%%%%%%%%
%%% SUMMARY %%%
%%%%%%%%%%%%%%%

In summary, we have investigated the influence of the cap layer
material and thickness on the thermal stability of bottom-pinned
spin-valves. Oxygen diffusion through the cap layer reduces the
GMR effect when the oxidation front reaches the CoFe free layer.
This can be avoided effectively by using a TaN cap. A second
diffusion mechanism leads to a continuous degradation of the GMR
performance. Co interdiffusion results in a loss  of Co in the
functional core region and an increased orange-peel coupling.
However, this detrimental effect occurs on a temperature and time
scale which is not relevant for the required thermal stability of
GMR devices in automotive applications.

This work was supported by the BMBF project 13N9084. The authors
would like to thank Pascal Verrier, Jacques Liebault and coworkers
of Altis Semicondutor, Corbeil Essonnes France, for the support in
deposition and evaluation of the GMR wafers.

\end{document}